\documentclass[floatfix,onecolumn,nofootinbib,longbibliography,preprintnumbers,10pt]{revtex4}

\usepackage{palatino}
\usepackage{makeidx}
\makeindex

\usepackage{braket} 

\usepackage{graphicx}

\usepackage{subfigure}

\usepackage{amssymb,amsmath,amsthm}

\newtheorem{postulate}{Postulate}

\theoremstyle{remark}
\newtheorem{remark}{Remark}

\theoremstyle{remark}
\newtheorem{example}{Example}

\theoremstyle{remark}
\newtheorem{consequence}{Consequence}

\usepackage{bm} 
\usepackage{graphicx} 
\usepackage{multirow} 
\usepackage{cases} 

\usepackage{booktabs,fancybox}

\usepackage{verbatim}  

\usepackage{xspace}
\makeatletter
\DeclareRobustCommand\onedot{\futurelet\@let@token\@onedot}
\def\@onedot{\ifx\@let@token.\else.\null\fi\xspace}

\def\eg{\emph{e.g}\onedot} 
\def\ie{\emph{i.e}\onedot} 
 
\def\etc{\emph{etc}\onedot}

\def\viz{\emph{viz}\onedot}
\makeatother

\usepackage{etoolbox} 
\newtoggle{omitauthor}
\newtoggle{omitwordcount}
\toggletrue{omitwordcount}
\newtoggle{omitdraftnumber}
\toggletrue{omitdraftnumber}
\newtoggle{omitdraftfootnote}
\toggletrue{omitdraftfootnote}

\usepackage[shortlabels,inline]{enumitem}  
\newlist{inlinelist}{enumerate*}{1}  
\setlist[inlinelist]{label=(\roman*)}  
\usepackage{hyperref}

\usepackage{caption}
\usepackage{subcaption}  

\usepackage{chngcntr}

\newcommand{\plus}{\ensuremath{\text{+}}}
\newcommand{\minus}{\ensuremath{\text{--}}}

\newcommand{\numberfield}[1]{\ensuremath{\mathbb{#1}}} 

\newcommand{\vect}[1] {\ensuremath{\mathbf{#1}}} 

\newcommand{\evol}[1]{\check{#1}}  

\newcommand{\tableheadtext}[1]{\textsf{\textbf{\footnotesize #1}}} 
\newcommand{\tabletext}[1]{\footnotesize \textrm{#1}} 

\newcommand{\op}[1]{\mathsf{#1}}   

\newcommand{\property}[1]{\ensuremath{\textsf{prop}(#1)}}

\newcommand{\opset}[1] {\ensuremath{\mathcal{#1}}} 

\newcommand{\mment}[1]{\ensuremath{\mathbf{#1}}} 

\newcommand{\seq}[2]{\ensuremath{[#1,#2]}}   

\newcommand{\lseq}[3]{\ensuremath{[#1,#2,#3]}}   

\newcommand{\llseq}[4]{\ensuremath{[#1,#2,#3,#4]}}   

\newcommand{\prob}[1]{\ensuremath{P(\sseq{#1})}}    
 
\newcommand{\sseq}[1]{\ensuremath{#1}}    

\newcommand{\bubble}[2]{\ensuremath\{#1,#2\}}   

\newcommand{\pll}{\lor}  

\newcommand{\ser}{\ensuremath{\cdot\,}}  

\newcommand{\amp}{z}  

\newcommand{\system}[1]{\ensuremath{\textsf{#1}}} 

\begin{document}

\title{From Complementarity to Quantum Properties: \\ An Operational Reconstructive Approach\iftoggle{omitdraftfootnote}{}{\footnote{This is a draft.  Comments, suggestions, and critical feedback are most welcome.}}}

\iftoggle{omitauthor}{}
{\author{Philip Goyal}	
    \email{pgoyal@albany.edu}
    \affiliation{University at Albany~(SUNY), NY, USA}
}
\date{\today}
\iftoggle{omitauthor}{}{\homepage[Homepage:~]{https://www.philipgoyal.org}}
\linespread{1.418}

\begin{abstract}
Quantum theory brings into question the compatibility of the twin desiderata of \emph{exact knowability} of the present state of the physical world and \emph{perfect predictability} of its future states.  Bohr's coordination--causality complementarity principle transforms this tension into one between properties~(as ordinarily understood in classical physics) and deterministic causality.  Here, we develop an explicit model of quantum properties which accommodates this essential tension.  Our approach integrates \emph{operational}, \emph{reconstructive}, and \emph{metaphysical} standpoints.  In particular, we make use of an operational framework employed in a recent operational reconstruction of Feynman's formulation of quantum theory; base our property model on an analysis of property types; and use the notions of actuality and potentiality to frame the model.  We show that this quantum property model provides a natural resolution of Zeno's paradox of motion, and provides reliable intuitions about phenomena such as electron diffraction and the non-local behaviour of entangled states of non-identical particles.
\end{abstract}
\maketitle

\section{Introduction}
Classical physics is shaped by the twin desiderata of \emph{exact knowability} of the present state of the physical world and \emph{perfect predictability} of its future states.  
However, quantum theory has brought the ultimate compatibility of these desiderata into question.  The clearest, albeit informal, expression of this incompatibility is \index{Bohr!Complementarity!Coordination--Causality Complementarity} Bohr's principle of \emph{coordination--causality complementarity}, which posits that exact space-time coordination of a physical object precludes the applicability  of the concept of causality~(as embodied, for example, in the principle of the conservation of momentum) to its behaviour\footnote{``The very nature of quantum theory thus forces us to regard the space-time coordination and the claim of causality, the union of which characterizes the classical theories, as complementary but exclusive features of the description [...].''~\cite[p.~580, para.~3]{Bohr1928}.}.

According to Bohr, this incompatibility originates in the failure of the classical ideal of a detached instrument of observation, \viz in the non-existence of a process of observation which simply registers the space-time location of an object as precisely as desired without affecting it in any way.
Yet, although these ideals~(exact space-time coordination; deterministic causality) are strictly incompatible, the complementarity principle asserts that they each provide an essential \emph{conceptual perspective} on microphysical reality, and that these perspectives are \emph{jointly necessary} for an understanding of microphysical phenomena\footnote{Bohr formulated complementarity in several different ways over a span of approximately thirty years, beginning with~\cite{Bohr1928}.  The main distinction between them is whether a formulation concerns \emph{different conceptual perspectives} on a single physical situation~(\emph{Complementarity I}, formulated \index{Bohr!Complementarity!Wave-Particle Complementarity@Wave--Particle Complementarity}\eg as wave--particle or coordination--causality complementarity) or concerns what can be learned about the same microphysical object via \emph{different experimental arrangements}~(\emph{Complementarity II}).  According to~\cite{Held1994}, Bohr's engagement with the EPR paper in 1935 marks the watershed, after which he emphasized Complementarity~II.  Here we focus on Complementarity I, which we regard as the farthest reaching formulation of the notion of complementarity~\cite[\S4]{Goyal2019a}.  Indeed, we here show that a precise complementarity principle of type~I can be extracted from an operational reconstruction of quantum theory.}.

Bohr's complementarity principle raises a number of key questions:
\begin{enumerate}
\item \emph{Origin of incompatibility of ideals.} Are there fundamental, \emph{a priori} reasons for anticipating that the above-mentioned ideals are incompatible? 
\item \emph{Origin of complementary perspectives.} Can the two above-mentioned conceptual perspectives, and their synthesis, be systematically derived from fundamental notions?  
\item \emph{Quantum properties.}  Can one speak of the properties of a quantum object in a manner which differs from classical physics, yet is grounded in the quantum formalism and provides empirically-robust, non-trivial intuitions about the behaviour of quantum systems?
\end{enumerate}
\emph{To elaborate:} Bohr's complementarity principle arose historically through an effort to interpret de Broglie's wave-particle duality.  And, following the formulation of Heisenberg's uncertainty relation, the notion of complementarity became a central pillar of Bohr's interpretation of quantum theory.  However, the plausibility of the principle rests primarily on the known behaviour of light and matter~(as, for instance, employed in Heisenberg's analysis of a light microscope~\cite{Heisenberg27}).  Accordingly, a deeper understanding of the principle~(specifically of its asserted incompatibility of ideals and complementarity of perspectives) calls for an analysis that proceeds from a deeper foundation, ideally one that does not invoke any complex phenomena~(such as electron diffraction) characteristic of quantum physics. 
 
In addition, according to Bohr's principle, one cannot meaningfully speak of the exact location \emph{and} the exact state of motion of a physical object as in classical physics.  Thus, Bohr's principle is \emph{restrictive} rather than \emph{prescriptive}---it restricts the validity of classical concepts~(in particular causality and property), but does not prescribe any definite alternative to the classical formulation of these concepts.  Now, one might say that the quantum formalism itself embodies Bohr's complementarity, and the mathematical language of quantum states and measurement observables obviates any need for a new property concept fitted to quantum objects.  However, while the property concept is largely avoidable at an instrumental level, quantum theory offers the profound opportunity to evolve our fundamental \emph{metaphysical} notions, such as substance and property~(which form a vital part of the metaphysical foundations of classical physics), in light of a battle-tested physical theory.  A precisely-defined notion of quantum properties would be an important step in this direction.

Since the formulation of quantum theory in the mid-1920s, a number of proposals have been put forward for applying the property concept to quantum systems.  For example, it is commonly asserted and widely accepted that if a quantum system is in an eigenstate of a measurement operator, then it possesses a property-value of the property corresponding to the operator.  However, the question of what can be said in the case where the system is not in an eigenstate of a given operator has received a wide range of answers.  For example, numerous authors, beginning with Heisenberg~\cite{Heisenberg1958} and Margenau~\cite{Margenau1954} in the mid-1950s, proposed that the state be regarded as encoding objective potentialities, an idea that is reflected in many subsequent works~(\eg~\cite{Aerts2010,Heelan2016, Shimony1997, Suarez2004,Karakostas2007, Jaeger2017, Kastner2018}).

Previous attempts to develop an understanding of quantum properties invariably employ the prevailing interpretative methodology, namely that the quantum formalism is taken as the basis of conceptual and philosophical analysis.  However, this methodology is inherently limited~\cite[\S2]{Goyal2025a}, for two reasons.  First, it tends to focus on features of the formalism whose meaning which can be readily grasped, neglecting those features---such as the formalism's mathematical structures---that are relatively opaque to philosophical scrutiny.  Second, it generally ignores the extensive \emph{know-how}---the modelling heuristics and experimental procedures---that is needed to connect the abstract quantum formalism to the sensory realm, despite the fact that this know-how invokes---usually implicitly---highly non-trivial assumptions that underpin quantum theory's claim to empirical adequacy.   However, the emergence of detailed principled derivations---or \emph{reconstructions}---of the quantum formalism over the past twenty-five years offers the basis for a new interpretative methodology~\cite[\S3]{Goyal2025a} which is free of these limitations.

In this paper, we approach the challenge of building a model of quantum properties by harnessing a recent operational \emph{reconstruction} of the quantum formalism~\cite{GKS-PRA, Goyal2014}.   We begin by establishing a detailed operational framework which enables  the above-stated questions to be approached in a manner that is relatively theory-agnostic.  We then switch to a metaphysical stance, referring as needed to the assumptions made in the reconstruction, to synthesize a quantum property concept that is manifestly a controlled generalization of the classical property concept.  A summary of the development of the quantum property model follows below.

\paragraph{Operationality.} \index{Operationalism} In classical physics, properties such as position and velocity are regarded as the `objective possession' of a physical object~(although their \emph{values} may considered to be relative to a frame of reference).  This reflects the metaphysical foundations of classical physics, whose assumptions license assertions about `what the physical world is really like in itself', independent of an agent's interests and purposes.  However, from an operational standpoint, the notions of persistent object and its properties are simply \emph{conceptual devices} for constructing \emph{predictive models} of \emph{appearances}.  In other words, the idea that there exist persistent objects and the idea that these objects bear specific properties are \emph{not free-standing facts-of-the-matter}, but are to be primarily understood in the context of the definite goal of \emph{prediction}.  And since we are obliged to employ the notion of causality~(even if in attenuated form) in order to build a predictive model that threads appearances across time, the notion of property is~(from this standpoint) secondary to the notion of causality.  

Here we use the operational framework described in~\cite[\S\,II\,A]{Goyal-QT2c}~(and subsequently elaborated in~\cite[\S\,II]{GKS-PRA}) to \emph{operationalize the property concept.}  The key organizing idea of the framework is \emph{causal closure}, which is used to define a set of measurements and transformations that probe the same subsystem of a physical object.  Using this framework, we systematically introduce properties whilst adhering to the principle of parsimony by requiring that \emph{a property is assigned if and only if it has predictive import.}

\paragraph{Reconstruction.}  \index{Reconstruction of Quantum Theory} Most interpretative accounts of quantum theory take the mathematical formalism of the theory as a starting point, and attempt to `read off' meaning from the formalism~(possibly after reformulation).  However, such an approach is inherently limited since, unlike the theories of classical physics, very little of the physical content of quantum theory can be `read off' the formalism of the theory---most of its rich physical content is encapsulated in its abstract mathematical structures, objects, and operators~(complex vector spaces, unitary operators, tensor products, \etc)~\cite{Goyal2025a}.  However, over the last two decades, the program of \emph{quantum reconstruction} has shown that the quantum formalism can be distilled into a small number of postulates which are more amenable to conceptual analysis~\cite{Fuchs2003,Grinbaum-reconstruction-interpretation,Goyal2022c,Berghofer2023,Goyal2025a}, thereby making its rich physical content more readily available for interpretation.

Here, we interpret a reconstruction of the Feynman rules of quantum theory~\cite{GKS-PRA, Goyal2014} to formulate a new \emph{property complementarity} principle.  This principle posits that, after a measurement yields an outcome, one can make two different property ascriptions to the object.  These property-models agree in their property ascriptions if the outcome is atomic~(\eg a point in space), but disagree if the outcome is non-atomic~(such as a spatial region,~$R$, corresponding operationally to the field of sensitivity of a detector).  The principle further posits that these property ascriptions are \emph{jointly} necessary for a property-based understanding of quantum objects, a posit that mirrors---and is thus justified by---the corresponding mathematical synthesis that takes place in the reconstruction.

\paragraph{Metaphysics.}  Finally, switching to a metaphysical stance, we argue that the complementary property models are naturally and appropriately synthesized using Aristotle's notions of actuality and potentiality.  For example, immediately after a non-atomic  outcome corresponding to spatial region~$R$, one can say that the quantum object is at region~$R$ \emph{actually}~(\ie the object is here modelled as an \emph{extended simple}) but is at any subregion~$R' \subset R$~(or point~$p \in R$) \emph{potentially.}  We show that this quantum property concept provides a natural resolution of Zeno's paradox of motion~(which historically was one of a family of paradoxes that provoked Aristotle's notions of potentiality and actuality), and provides reliable intuitions about phenomena such as electron diffraction and the non-local behaviour of entangled states of non-identical particles.

The paper is structured as follows.  In Sec.~\ref{sec:operational-framework}, we present the operational framework.  In Sec.~\ref{sec:properties}, we analyse property types, focussing on instantaneous categorical properties; and then operationalize classical and quantum properties.  In Sec.~\ref{sec:complementarity}, we formulate the principle of property complementarity on the basis of the above-mentioned reconstruction.  In Sec.~\ref{sec:actual-potential-quantum-properties}, we reframe property complementarity in terms of the notions of actuality and potentiality.  In Sec.~\ref{sec:examples}, we show how the proposed model of quantum properties offers a novel resolution of Zeno's arrow paradox and provides reliable intuitions about double-slit diffraction and non-local behaviour in systems of entangled non-identical particles.  In Sec.~\ref{sec:related-work}, we situate the present work in the broader context of some related work.  We conclude in Sec.~\ref{sec:discussion} with a brief discussion, including some open questions.

\section{Operational framework}
\label{sec:operational-framework}

In order to systematically compare classical and quantum theories side-by-side, it is invaluable to construct an \emph{operational framework} within which each can be embedded.  Such a framework minimizes the explicit or implicit appeal to theoretically-laden notions~(such as property) whose meaning and role can differ markedly between theories, and brackets metaphysical presuppositions that accompany such notions.  Instead, the goal is to precisely describe an ideal experiment on a physical system, relying on primitive notions such as \emph{measurement}, \emph{measurement outcome}, and \emph{interaction}; and on mathematical tools such as probability theory, which are equally applicable to different physical theories.  The operational framework described here is based on that described in~\cite[\S\,II\,A]{Goyal-QT2c} and subsequently elaborated in~\cite[\S\,II]{GKS-PRA}.  Although this operational framework was employed in~\cite{Goyal-QT2c,GKS-PRA} as the basis for reconstructions of the quantum formalism, it is presented here in a theory-agnostic manner.

\subsection{An Ideal Experiment and its Theoretical Modelling}
\label{sec:ideal-experiment-and-modelling}

The operational framework rests on the idea that a \emph{physical system} is subject to an \emph{experiment}, namely a sequence of \emph{measurements} and \emph{interactions}.   The measurements and interactions are here conceived abstractly as physical processes that can act on the system without it losing its identity.  Measurements differ from interactions in that measurements yield \emph{outcomes}.  Such outcomes are ordinarily manifest as macroscopically-observable events~(such as flashes on a scintillation screen), but here they are abstractly conceived as classically-definite events.  The measurements are restricted to those that are \emph{repeatable}---\viz if a measurement performed at~$t$ yields outcome~$m$, then immediate repetition of the measurement yields the same result with certainty\footnote{\label{fn:repeatability}A more precise statement of repeatability for measurements with countably many outcomes is as follows.  Suppose that the measurement is performed at~$t$ and yields outcome~$m$.  Then, for any~$m$, repetition of the measurement at~$t + \delta t$ yields outcome~$m$ with probability~$p$ that tends to unity as~$\delta t \to 0$.  The statement for the uncountable case is similar.  Suppose the measurement at~$t$ yields outcome~$\vect{r} \in \numberfield{R}^n$, and suppose the same measurement at~$t + \delta t$ has probability distribution~$p(\vect{r}')$ over its possible outcomes~$\vect{r}'$.  Then, for any~$\vect{r}$, we require that, for any~$\epsilon > 0$, the probability contained within a ball of radius~$\epsilon$ around~$\vect{r}$ tends to unity as~$\delta t \to 0$.}.  We emphasize that, at this stage, \emph{the notion of property is entirely absent}---a measurement is not conceived as a measurement \emph{of} a property, nor as preparing a system with some value of a property.  The property notion will be systematically introduced in Sec.~\ref{sec:property-operationalization}.  

The operational framework is itself shaped by the goal of constructing a theoretical model that is capable of predicting the probabilities of the possible outcomes of a measurement in such an experiment.  This leads to the ideal of an experiment which~(in a precise sense to be discussed below) isolates the system under investigation from its past history insofar as the aspects of the system scrutinized by the experiment are concerned.

\subsection{Defining an Experiment via Causal Closure}

It is common to say that an experiment is performed on a \emph{particle} or a \emph{spin}.  However, such notions refer to idealized physical systems rather to the physical objects given to us in nature~(insofar as any object is given to us pre-individuated).  For example, viewed from a classical atomistic perspective, a billiard ball is a complex object with many degrees of freedom.  It can be struck by a hammer or other tool, and thereby made to ring in a complex manner; it can be physically deformed in myriad ways; its surface can be roughened and (re)painted.  Thus, an experiment `on a billiard ball' must be carefully circumscribed if it is to yield useful information.  In essence, an ideal experiment must isolate specific degrees of freedom for the duration of the experiment, and must prepare the relevant degrees of freedom with known values at the outset.  Such initial preparation renders the pre-preparation history of the object irrelevant insofar as the results of the experiment are concerned.  Although such preparations may not always be possible, they are the ideal against which an experiment with non-ideal preparation is gauged.   At this point, the notion of \emph{causality} has entered:~we shall say that an ideal preparation establishes \emph{causal closure} with respect to subsequent measurements~\cite[\S2A]{Goyal-QT2c}.

Consider, for example, an experiment which is designed to probe the centre-of-mass spatial behaviour of a classically-describable billiard ball.  A preparation would consist in fixing the initial position and velocity of the ball.  Subsequent measurements would need to be restricted to those which only probe these degrees of freedom~(position, velocity).  For example, a measurement whose outcome depends upon the ball's rotational motion or, say, its material composition or colour, would need to be excluded.  Similarly,  interactions with the ball would need to be restricted to those that do not couple its spatial~(position, velocity) degrees of freedom to its rotational or non-spatial degrees of freedom.  For instance, interactions which change the ball's velocity in a way that depends upon its colour would need to be excluded.  With these restrictions in place, the outcomes of measurements on the ball would be independent of pre-preparation interactions with it~(such as re-painting it a different colour), and the experiment would probe only the spatial sub-component of the actual physical object.  One could then describe the experiment as one in which the billiard ball is being treated \emph{as if} it were a particle~(\viz an abstract object whose only time-dependent properties are position and velocity); equally, one could say that the experiment probes the particle subsystem of the object.   

\subsubsection{Closure condition.}  Experimental design usually appeals to a pre-existing theoretical model of the object.  In the above example, the design process appeals to a classical theoretical model that invokes the theoretical notions of \emph{physical state} and \emph{properties}.  However, in the operational framework, these notions are unavailable.  Hence, we must formalize the notion of causal closure and be able to pick out a subsystem without relying on such a pre-existing model.   

Consider, then, an experiment in which a physical system is subject to a \emph{preparation} followed by a  \emph{measurement}, with no interactions in between.  Suppose that the preparation at~$t_1$ is implemented by performing a measurement,~$\mment{M}$, and selecting the system that yields outcome~$m$; and that measurement~$\mment{N}$ is subsequently performed at~$t_2$, yielding outcome~$n$.  If the outcome probabilities,~$\Pr(n \,|\, m; M, N)$, of~$\mment{N}$ are independent of pre-preparation interactions with the system, then~$\mment{M}$ is said to \emph{establish closure} with respect to~$\mment{N}$.  If closure  also holds when~$\mment{M}$ and~$\mment{N}$ are interchanged, we say that~$\mment{M}, \mment{N}$ form a \emph{measurement pair}, and that they probe the same \emph{subsystem} of the system.  

Using this closure condition, we can define the set~$\opset{M}$ of all measurements which form a measurement pair with~$\mment{M}$ and are not composites of other measurements in~$\opset{M}$.  Similarly, the set~$\opset{I}$ consists of all interactions which are \emph{compatible} with~$\opset{M}$ in the sense that any interaction~$I \in \opset{I}$ can be performed between two measurements~$\mment{M}, \mment{M}' \in \opset{M}$ without disrupting closure.  Thus, by employing the closure condition, one can define the set of all measurements and interactions which probe the same subsystem of a system.

\paragraph*{Example: Stern-Gerlach experiment.}  Consider a Stern-Gerlach measurement performed on silver atoms.  A standard Stern-Gerlach device yields one of two possible outcomes.  A preparation implemented by such a Stern-Gerlach measurement establishes closure with respect to a subsequent Stern-Gerlach measurement.  Hence, the set of all such Stern-Gerlach measurements constitutes a measurement set,~$\opset{M} = \{\textsf{SG}_{\theta, \phi} \,|\, \theta \in [0, \pi), \phi \in [0, 2\pi) \}$.  In addition, all interactions consisting of the application of a uniform magnetic field between any two such measurements maintain closure.  However, interactions involving a non-uniform magnetic field are prohibited since they fail to maintain closure.

\subsection{Atomic and Non-atomic Measurement Outcomes}

In classical physics, an idealized position measurement yields a point-valued outcome~$\vect{r} \in \numberfield{R}^3$.  However, in practice, such an ideal is unattainable---one is limited to measurements with  detectors whose range of sensitivity is a finite spatial region, and the firing of such a detector corresponds to a region-valued outcome~$R \subset \numberfield{R}^3$.  In classical physics, if a region-valued outcome~$R$ is obtained, one would infer that the object is at some point~$\vect{r} \in R$.  Hence, one speaks of such a measurement as an \emph{inexact position measurement}.  However, such an inference is theoretically-laden---it depends upon a specific understanding of the relation between observational data and an object's properties.  Hence, in the operational framework, we shall describe such idealized and non-idealized measurements in a neutral manner.

In the operational framework, an outcome of a measurement is either \emph{atomic} or \emph{non-atomic.}  For example, the point-valued outcome of an idealized position measurement is atomic, while a region-valued outcome of a non-ideal position measurement is non-atomic.  If all of the possible outcomes of a measurement are atomic, we say the measurement is atomic; otherwise it is non-atomic.

Abstractly, an outcome~$m$ of measurement~$\mment{M}$ is atomic if it cannot be \emph{refined}.  That is, there is no measurement which, when performed immediately afterwards, will yield more than one possible outcome with non-zero probability if~$\mment{M}$ yields~$m$, but yield only one outcome with certainty otherwise.  Conversely, outcome~$\widetilde{m}$ is non-atomic if \begin{inlinelist} \item  it can be refined; and \item the repeatability of any refinement,~$m_i$ of~$\widetilde{m}$ is unaffected by~$\widetilde{m}$ being obtained in between the repetition. \end{inlinelist}  If the refinements of~$\widetilde{m}$ consists in the atomic outcomes~$m_1, \dots, m_k$, we shall write~$\widetilde{m} = \{m_1, \dots, m_k\}$.


\paragraph*{Example: Atomic and non-atomic Stern-Gerlach measurements.} A standard Stern-Gerlach measurement performed on silver atoms yields one of two possible outcomes, which we can denote~$1$ and~$2$.  These are atomic outcomes since they cannot be refined through further measurements.  We shall accordingly refer to the detectors themselves as atomic.  Conversely, if the atomic detectors are coarse-grained to yield a single detector, the resulting measurement has only one possible outcome.  Since this outcome can be refined by a standard Stern-Gerlach measurement, it is non-atomic and can be denoted~$\bubble{1}{2}$.

\subsection{Sequence Probabilities and their Interrelations}
\label{sec:sequence-probabilities-and-interrelations}
\begin{figure}[!h]
\centering
\subfigure[\,Stern-Gerlach experiment with a sequence of atomic outcomes.]
				{
				\includegraphics[width=4.5in]{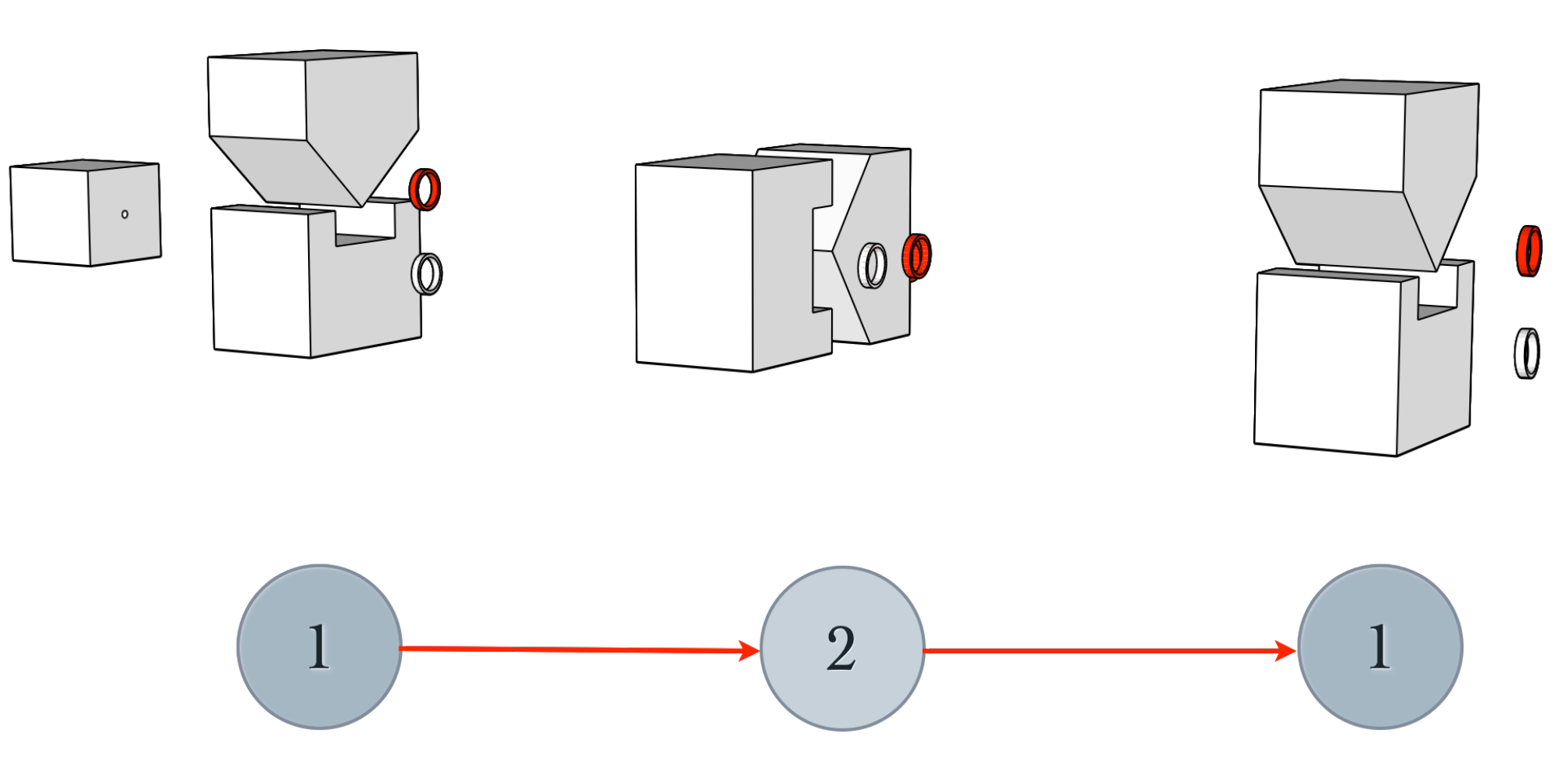}
				\label{fig:SG-experiment}
				}
\subfigure[\,Stern-Gerlach experiment with a coarse-grained outcome sandwiched by two atomic outcomes.]
				{
				\includegraphics[width=4.5in]{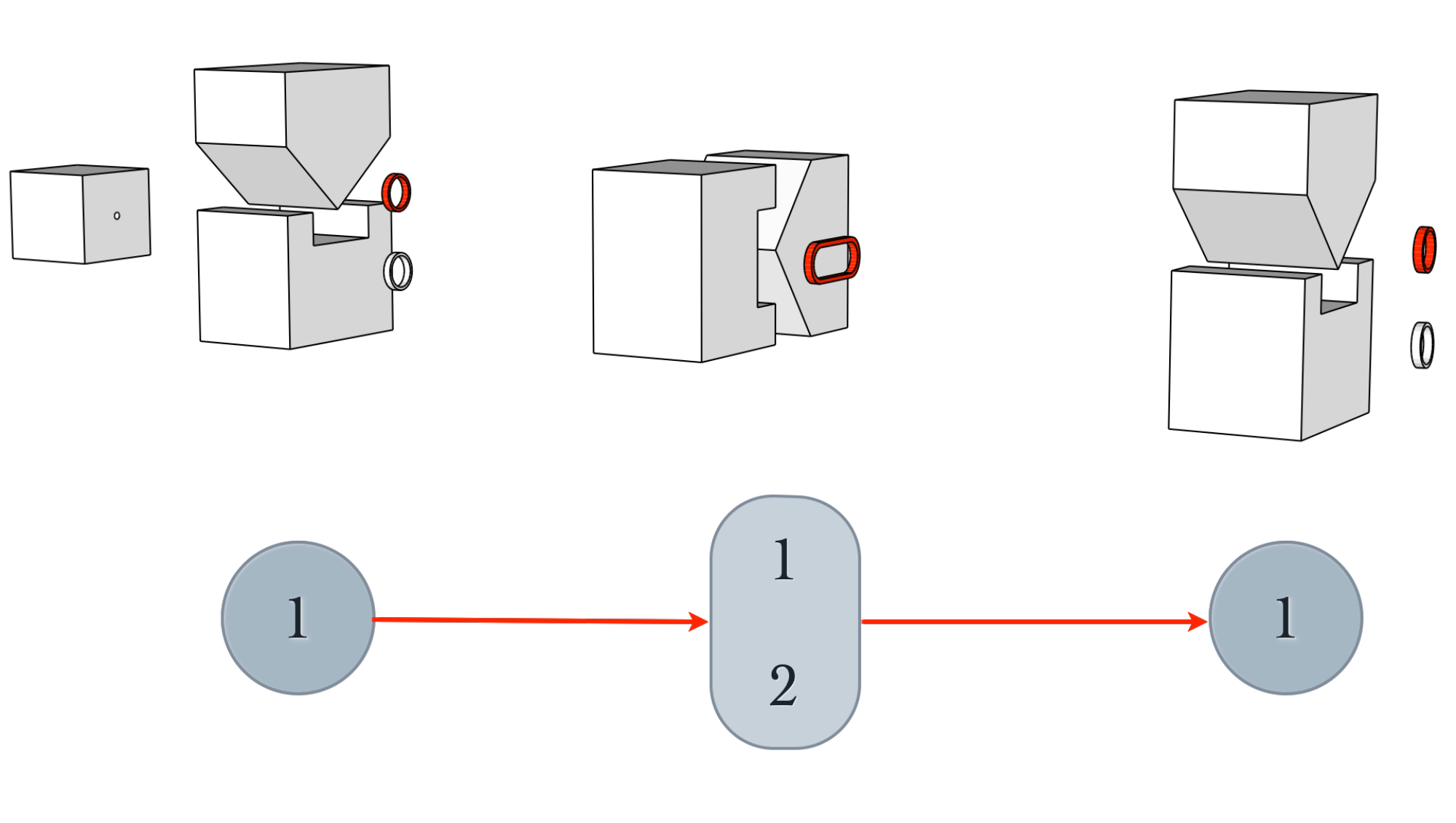}
				\label{fig:SG-experiment-coarse}
				}
\caption[]{\label{fig:Stern-Gerlach-experiment} Stern-Gerlach experiments on silver atoms.  In~(a), a silver atom is subject to a sequence of atomic Stern-Gerlach measurement.  Each measurement yields one of two atomic outcomes, labelled~$1$ or~$2$.  The sequence of outcomes is denoted~$\lseq{1}{2}{1}$.  In~(b), the intermediate Stern-Gerlach measurement yields non-atomic outcome~$\bubble{1}{2}$.  The sequence of outcomes is denoted~$\lseq{1}{\bubble{1}{2}}{1}$.  Using Eq.~\eqref{eqn:parallel-combination}, we can regard this sequence as the combination~$\lseq{1}{1}{1} \pll \lseq{1}{2}{1}$.}
\end{figure}

Consider an experimental set-up in which a physical system is subject to successive measurements~$\mment{L}, \mment{M}, \mment{N}$ at successive times~$t_1, t_2, t_3$, with there possibly being interactions with the system in the intervals between those measurements~(see example in Fig.~\ref{fig:SG-experiment}).  The measurements and interactions in any such set-up are chosen in accordance with the closure condition described above.  The outcomes obtained in a given run of the experiment are summarized in the \emph{sequence}~$\sseq{C} = \lseq{\ell}{m}{n}$.  Due to closure, outcome~$\ell$ is atomic.

In general, the probability~$\prob{A}$ associated with sequence~$\sseq{A} = \llseq{\ell}{m}{n}{\dots}$ is defined as the probability of obtaining outcomes~$m, n, \dots$ conditional upon obtaining~$\ell$,
\begin{equation} \label{eqn:def-of-probability-of-sequence}
\prob{A} = \Pr(m, n, \dots \,|\, \ell).
\end{equation}
Since closure has been established by outcome~$\ell$ of measurement~$\mment{L}$, the sequence probability~$\prob{A}$ is independent of interactions with the system prior to~$t_1$, and hence is well-defined.

Now, a predictive model establishes interrelations between the probabilities of different sequences.  To set the stage for such interrelations, we introduce some formalism which expresses the idea that experimental set-ups are interrelated in particular ways.   We focus here on two specific set-up interrelations; others are introduced in~\cite{GKS-PRA,Goyal2014}.

First, we can view the above set-up as a \emph{series concatenation} of two experiments, the first in which measurements~$\mment{L}$ and~$\mment{M}$ occur at times~$t_1$ and~$t_2$, yielding the sequence~$\sseq{A} = \seq{\ell}{m}$, and the second in which measurements~$\mment{M}$ and~$\mment{N}$ occur at times~$t_2$ and~$t_3$, yielding~$\sseq{B} = \seq{m}{n}$.  In order to ensure that closure is satisfied in the second experiment, measurement~$\mment{M}$ must be atomic.  Formally, 
\begin{equation}
\sseq{C} = \sseq{A} \ser \sseq{B},
\end{equation}
where~$\ser$ is the \emph{series} combination operator.  
Under these conditions, one finds~\cite[\S\,III\,B\,1]{GKS-PRA} that the probability~$\prob{C}$ is related to~$\prob{A}$ and~$\prob{B}$ via
\begin{equation}
\label{eqn:probability-product}
\prob{C} = \prob{A} \cdot \prob{B}.
\end{equation}

Second, one can consider a set-up which is identical to the one above, except that outcomes~$m$ and~$m'$ of~$\mment{M}$ have been coarse-grained, so that one obtains the sequence~$\sseq{E} = \lseq{\ell}{\bubble{m}{m'}}{n}$.  Formally, we express the relationship of this sequence to the sequences~$\sseq{C} = \lseq{\ell}{m}{n}$ and~$\sseq{D} = \lseq{\ell}{m'}{n}$ as
\begin{equation}
\label{eqn:parallel-combination}
\sseq{E} = \sseq{C} \pll \sseq{D},
\end{equation}
where~$\pll$ is the \emph{parallel} combination operation.  
At this stage, there is no necessary relationship between the probabilities~$\prob{C}, \prob{D}$ and~$\prob{E}$.  Hence, fixing the form of this relationship requires additional theoretical input.  As we shall discuss in Sec.~\ref{sec:complementarity}, in a classical probabilistic model, the relation is
\begin{equation}
\label{eqn:classical-parallel-probability}
P(E) = P(C) + P(D).
\end{equation}
However, in Feynman's rules of quantum theory~(when expressed in terms of the operational framework above), another relationship takes its place.

\section{Properties}
\label{sec:properties}

\subsection{Properties in Human Language and their Metaphysical Classification}
\label{sec:natural-properties-and-metaphysical-classification}
Human natural language has co-evolved with an informal yet sophisticated model of the physical world.  Central to this model is the view that the physical world contains physical objects~(whether animate or inanimate) that are instantiations of various types~(\eg `tree', 'person'), and that these objects possess various properties.

The metaphysical analysis of the property concept~(as manifest in natural language) draws a key distinction between \emph{categorical} and \emph{non-categorical} properties.  The former are idealized as being \emph{fully present} in the now:~to say that an object has such-and-such categorical property is to firmly assert \emph{what is} rather than to indicate \emph{what could be.}  Categorical properties include position and shape, which can be perceived in the now.  In contrast, non-categorical properties are not directly perceived, but are rather `indirectly perceived'~(or inferred) via categorical properties.  Furthermore, they are not manifested in the now but only over a \emph{duration}, and their manifestation is typically dependent upon certain conditions~(triggers).  Examples include \emph{tendencies} such as the fragility of a glass ornament, and \emph{capacities} such as a person's ability to sing.   As such, non-categorical properties can usually be regarded as ways of summarizing regularities abstracted from the past behaviour of physical objects~(whether living or non-living).  

The distinction between categorical and non-categorical properties raises the question of whether~(and in what sense) the former are more `real' than the latter, and whether the latter can always be reduced to the former.    As modern science is concerned with grounding knowledge on reproducible, intersubjectively-verifiable observations, the context-dependent element of many non-categorical properties is problematic.  Hence, non-categorical properties are generally restricted to those definable in terms of categorical properties in a context-independent manner, while behavioural regularities are generally factored out and formulated as explicit laws that refer only to categorical properties~(or to non-categorical properties that can be non-contextually defined in terms of them). 

\subsection{Properties in Classical Physics}
\label{sec:classical-properties-introduction}

Classical physics is founded upon the categorical properties of location and time, where (local)~time is in turn operationally defined in terms of location via an ideal clock.  Due to the real-valued representation of time, `the now' corresponds to a mathematical instant of time.  Thus precisified, location is an instantaneous categorical property~(ICP).  However, certain properties, such as velocity, are defined over an infinitesimal duration, and in this case a classification in terms of established metaphysical categories proves challenging.  Such so-called neighbourhood properties~\cite{Arntzenius2000} could be viewed as categorical~(\viz neighbourhood categorical properties~(NCPs)) insofar as they are regarded as deterministically manifest over an infinitesimal duration~(such a duration arguably better corresponding to our lived sense of `now' than a mathematical instant) or as non-categorical~(since \begin{inlinelist} \item they are not defined at a mathematical instant; and \item there are edge cases in which they may not be manifested \end{inlinelist})~\cite{Arntzenius2000,Lange2005,McCoy2018}.  In view of this ambiguity, we shall refer henceforth to properties such as velocity, which concern the infinitesimal change of ICPs, as \emph{evolutive} properties.  

\subsection{Properties in Quantum Theory}
\label{sec:quantum-properties-introduction}

As discussed in the Introduction, quantum theory brings into question the applicability of the notion of properties to describe physical objects.  To be sure, time-independent properties such as mass and charge are retained in quantum models of physical systems~(such as in the Schr\"odinger and Dirac equations).  However, the abstract quantum formalism~(\eg the von Neumann--Dirac axioms) make no reference to properties \emph{per se}.  Instead, the formalism refers only to \emph{observables}, represented by Hermitian operators.  A set of theoretical heuristics is then used to fix the mathematical form of the operators that correspond to certain concrete measurement procedures.  A particular challenge in establishing how quantum theory utilizes the property concept is that, by their nature, these heuristics are nowhere spelled out in detail.  Nevertheless, they are essential for the construction of quantum models of specific systems of interest.

A key assumption is that a measurement procedure that we classically recognize as implementing a position measurement~(\ie a measurement of the position property of an object) is also a valid measurement on a quantum system, and that the procedure quantum theoretically constitutes a `measurement of position' in a qualified sense.  The caveat here is that, in the quantum context, it is presumed only that such a measurement \emph{prepares} a quantum object in a quantum state that can be interpreted as one describing an object that has a definite spatial position.  In particular, it is not assumed that the outcome of the measurement procedure yields the value of the position of the quantum object immediately prior to measurement.  Indeed, due to its limited applicability, property language is generally eschewed, surviving only in the language used to describe the measurement operators~(`position operator', \etc).

However, the application of the quantum formalism favours \emph{specific} classical properties.  For example, while a position measurement procedure is assumed to be describable by a corresponding operator, the idealized classical procedure for measuring velocity or momentum is \emph{not}.  The experimental procedure that represents a measurement of momentum in the quantum context is obtained indirectly.  For example, one can employ classical-quantum correspondence arguments to derive a commutation relation between~$\hat{x}$ and~$\hat{p}_x$, and solve to obtain an explicit coordinate representation of~$\hat{p}_x$ given that~$\hat{x}$ is given by~$x$ in the coordinate representation of the quantum state of a `particle'~\cite{Goyal-correspondence-rules-2010}. Knowing~$\hat{p}_x$, one can then construct an experimental procedure which implements a measurement of~$\hat{p}_x$.  But such a procedure is \emph{completely unlike} the classical procedure for measuring momentum.  Hence, the sense in which~$\hat{p}_x$ is a `momentum observable' is \emph{substantially different} from the sense in which~$\hat{x}$ is a `position observable',  a fact that is obscured by conventional terminology.

The contrast between the way in which the quantum measurement operators for position and momentum are obtained, and the way that the corresponding measurement procedures are obtained, is striking.  In terms of the metaphysical distinctions introduced above, we can summarize the situation by saying that measurement procedures which, classically, are deemed to be measurements of ICPs appear to be the \emph{common ground} between classical and quantum physics.  That is, \emph{if a procedure in classical physics is taken to be a measurement of an ICP, then this \emph{same} procedure is taken to be a candidate measurement of an ICP in the quantum context}~(with the caveat that certain ICP measurements, such as angle measurements, may not have operator representations that satisfy certain minimal requirements expressible within the quantum formalism).

\subsection{Operationalization of Physical Properties}
\label{sec:property-operationalization} \index{Operationalism}

In order to place the attribution of properties to classical or quantum systems on a firmer conceptual footing, we shall now systematically introduce properties in an operational setting.  This entails a shift away from the traditional metaphysical view that properties are pre-given attributes of physical objects, to the view that properties are attributed to objects if and only if such attribution is justified in an operational setting.  Our broad objective is to finely \emph{differentiate} property attribution statements so that it is clear which specific operational facts or theoretical assumptions ground which property attributions.

As indicated in Sec.~\ref{sec:operational-framework}, the notion of closure is central to the operational framework.  The framework assumes that there exist measurements that establish closure, but offers no explicit guidance on how to \emph{find} these measurements.  As our discussion above~(Sec.~\ref{sec:quantum-properties-introduction}) suggests, however, measurements of ICPs are the common ground between classical and quantum mechanics.  Accordingly, below, we specify the closure conditions for these theories in terms of ICP measurements.

\subsubsection{Pre- and post-measurement property attribution}
\label{sec:repeatability-property-attribution}

\paragraph{Repeatability grounds property attribution.} \label{sec:repeatibility-property-attribution}
When a measurement is performed on a physical system and yields an outcome, immediate repetition of the measurement yields the same outcome with certainty~(see footnote~\ref{fn:repeatability} for a more precise statement).  This primitive operational fact grounds the intuition that, immediately after the first measurement, the object has been \emph{imprinted} in a precise manner that corresponds to the measurement outcome.  In turn, this intuition grounds the assertion that, immediately after the measurement, the object \emph{possesses} a property corresponding to the measurement outcome\footnote{In this paper, property ascriptions are always property-valued---\ie there is no case where a system is ascribed a property~(such as position) without a corresponding value.  Hence, the determinate--determinable distinction is not applicable.}.

\paragraph{Property attribution following atomic and non-atomic outcomes} \label{sec:atomic-nonatomic-property-attribution} \label{sec:property-attribution-non-atomic-outcomes}

\begin{enumerate}[(i)]
\item\emph{Atomic outcomes.}  Immediately after a measurement~$\mment{M}$ yields atomic outcome~$m$, one can attribute a corresponding property with value~$\property{m}$ to the object.  For example, immediately after an idealized position measurement\footnote{We assume throughout that ideal position measurements yield real-valued outcomes.  If spacetime is presumed to have a fundamentally discrete structure, this assumption would need to be modified.}  yields a flash at a point~$\vect{r}$ in space at~$t$, one can say that, at~$t^\plus \equiv t + dt$, the object \emph{is located} at~$\vect{r}$, \ie has position property with value~$\vect{r}$. 

\item\emph{Non-atomic outcomes.} After a measurement~$\mment{M}$ yields non-atomic outcome~$\widetilde{m}$, one can attribute a property of type corresponding to~$\mment{M}$ whose value is \emph{compatible} with~$\widetilde{m}$.  Without further theoretical constraints on the type of property-value~(\eg whether it is point-valued or whether it can be region-valued), a more precise statement cannot be made.  For example, after a non-ideal position measurement yields a region-valued outcome~$R \subset \numberfield{R}^3$, one can say that the object has a position property which is compatible with~$R$.  Absent further theoretical input, one cannot say that it has a \emph{point-valued} position property~(and is accordingly located some \emph{point} $\vect{r} \in R$).
\end{enumerate} 

\paragraph{Pre-measurement property attribution}
Measurement repeatability does not ground \emph{retrodictive} property attributions.  For example, if a position measurement yields outcome~$\vect{r}$, one cannot assert on the basis of repeatability that the object \emph{was} located in the vicinity of~$\vect{r}$ at~$t^\minus \equiv t - dt$.  Whether or not this assertion is correct will depend upon additional theoretical assumptions.   Hence, from the operational perspective, there is a marked distinction between pre- and post-measurement property-ascriptions.

\subsubsection{Operationalization of classical properties}
\label{sec:operationalization-classical-properties}
\paragraph{Classical closure.}
In classical mechanics, the state,~$S_1$, of a system of particles at~$t_1$ determines the state,~$S_2$, at~$t_2$ given the equations of motion and a description of the particle's environment in~$[t_1, t_2]$.  That is, given~$S_1$, the state~$S_2$ is independent of any interactions with the particles prior to~$t_1$.  This motivates the following \emph{classical closure} postulate~(CCP):
\begin{postulate}[Classical Closure] \label{postulate:instantaneous-property-pair-closure}
There exist pairs of instantaneous categorical properties~(ICPs)~$P, P'$, such that immediately successive atomic measurements of~$P, P'$, performed at times~$t, t+dt$, establishes closure with respect to a subsequent measurement~(atomic or non-atomic) of~$P$ and/or~$P'$.
\end{postulate}

\begin{example}[Closure for single classical particle]
A pair of immediately successive atomic measurements of a position performed upon a classical particle 
establishes closure with respect to subsequent position measurements.
\end{example}

\begin{example}[Closure for classical rotating triad~(`spin')]
A pair of immediately successive atomic measurements of the configuration of a classical spin establishes closure with respect to subsequent triad configuration measurements.
\end{example}

\paragraph{Registrative and passive measurements}
\label{sec:registrative-passive-measurements}
\begin{enumerate}[(i)]
\item\emph{Registrativity.}  Classical physics assumes that an ideal classical measurement is \emph{registrative}, \ie its outcome deterministically reflects~(at least in part) the pre-measurement state.  This allows one to go beyond the property attribution statements that follow from repeatability alone~(as discussed in Sec.~\ref{sec:repeatability-property-attribution}), enabling one to attribute a property to the object immediately \emph{prior} to the measurement:
\begin{itemize} 
\item\emph{Pre-measurement property attribution.}  If a property is attributed to an object immediately following a \emph{registrative} measurement upon it, then that object possessed that property immediately prior to the measurement.
\end{itemize}

\item\emph{Passivity.}  A registrative measurement\footnote{Here and elsewhere, as stipulated in Sec.~\ref{sec:ideal-experiment-and-modelling}, measurements are assumed to be repeatable unless explicitly stated otherwise.} leaves the measured property unchanged.  However, it may not leave unchanged components of the state other than this property.  We define as \emph{passive} a measurement that leaves the state \emph{wholly} unchanged.  Hence, an ideal classical measurement is passive and registrative.  
\end{enumerate}

\paragraph{Evolutive properties.}
Suppose that closure is established through immediately successive atomic registrative measurements of ICP~$P$~(as opposed to two different ICPs~$P$ and~$P'$); and that these measurements, performed at times~$t$ and~$t'$, yield values~$P(t)$ and~$P(t')$.  Due to repeatability, one can attribute property~$P(t)$ to the object at time~$t^\plus$; and, due to repeatability and registrativity, the property~$P(t')$ at~$t'^\minus$.

The cause of the change in property~$P$ between times~$t^\plus$ and~$t'^\minus$ can, at least in part, be attributed to an \emph{evolutive} property,~$\evol{P}$, at~$t^\plus$.  This leads to the following postulate:
\begin{postulate}[Classical Evolutive Properties] \label{postulate:classical-evolutives}
Following atomic passive registrative measurements of instantaneous categorical property~$P$ at~$t, t + dt$, the object possesses an evolutive property,~$\evol{P}$, with value~$dP/dt$.  This value is well-defined since, due to measurement repeatability, the values of the outcomes of the immediately successive $P$-measurements differ at most infinitesimally.
\end{postulate}
\begin{example}[Evolutive property of a classical particle]
In the case where~$P$ is position,~$\evol{P}$ is instantaneous velocity.  
\end{example}

\begin{consequence}[Non-preclusion of deterministic outcomes]\label{cons:determinism}
Suppose that a pair of immediately successive atomic passive registrative position measurements are performed on a single particle.  As this initial data yields exact information about both the particle's initial location \emph{and} information about its instantaneous state of motion, this initial data does not preclude deterministic prediction of the outcome of a subsequent position measurement.  (Whether prediction is \emph{actually} deterministic would then depend upon other factors, in particular whether the laws of motion are deterministic, and whether the environment of the system is deterministically describable.)  In contrast, lack of information about its initial location or its initial state of motion would preclude deterministic prediction.
\end{consequence}

\paragraph{Property attribution following non-atomic outcomes.} \label{sec:classical-non-atomic-property-attribution} As discussed in Sec.~\ref{sec:property-attribution-non-atomic-outcomes}, property attribution following a non-atomic outcome~$\widetilde{m}$ is theory-dependent.  Classical physics assumes that properties are real-valued, to which we shall refer as the \emph{property atomicity} assumption.  Hence, if a non-ideal position measurement yields a region outcome~$R \subset \numberfield{R}^3$, classical mechanics asserts that the object is, in fact, at some \emph{point}~$\vect{r} \in R$.

\subsubsection{Operationalization of quantum properties}

\paragraph{Quantum closure.}
As discussed in Sec.~\ref{sec:quantum-properties-introduction}, the abstract quantum formalism does not, in itself, specify \emph{which} properties~(understood, as per Sec.~\ref{sec:quantum-properties-introduction},  as classically instantiated through standard experimental procedures) are representable as non-degenerate `observables'.  However, conventional application of the abstract formalism deems specific properties such as position and spin along a direction~(implemented by Stern-Gerlach devices) as thus representable.  This choice is amply supported through extensive empirical success\footnote{However, it is important to keep in mind that that this \emph{is} a choice.  Indeed, it has been recently challenged~(see e.g.~\cite{Beige2025}).}.  For example, the quantum wave equations~(Schr\"odinger, Dirac) all rest on the choice of position as representable as such\footnote{This can be seen clearly in reconstructions of these equations~(see, for example~\cite{Goyal-correspondence-rules-2010}).}.  This empirically-supported choice, together with the metaphysical considerations of Sec.~\ref{sec:natural-properties-and-metaphysical-classification}, motivates the following postulate:
\begin{postulate}[Quantum Closure] \label{postulate:quantum-closure}
Atomic measurement of instantaneous categorical property~$P$ establishes closure with respect to a subsequent measurement~(atomic or non-atomic) of~$P$.
\end{postulate}

\begin{example}[Atomic position measurement establishes closure] 
If an atomic position measurement is performed on a quantum object at time~$t$, then the outcome probabilities of any subsequent position measurement performed on the object are independent of interactions with the object prior to~$t$.
\end{example}

\paragraph{Evolutive properties.}
Postulate~\ref{postulate:quantum-closure} implies that, in experiment where an atomic measurement of an instantaneous property,~$P$, is performed at~$t$, the history of the system prior to~$t$ is irrelevant to the question `why is the outcome of a subsequent $P$-measurement~$p$?' Hence, we have \emph{no empirical warrant} to posit that the object possesses an evolutive property,~$\evol{P}$, at~$t$.  This leads to the following postulate:
\begin{postulate}[Quantum Property Exclusivity] 
If a quantum object possesses an exact value of an instantaneous property,~$P$, it possesses no corresponding evolutive property,~$\evol{P}$.
\end{postulate}

\begin{example}[Quantum particle] 
If a quantum object is exactly located at time~$t$, it possesses no evolutive property analogous to the instantaneous velocity or momentum of a classical particle.
\end{example}

\begin{remark}[Bohr's Coordination--Causality Complementarity] \label{rem:Bohr-coordination-causality}
The quantum particle example above shows that Quantum Property Exclusivity is a precise expression of the limiting case of Bohr's coordination--causality complementarity.  That is, in the limit of precise space-time coordination, the object possesses no evolutive property, and hence the notion of a conservation principle~(which is formulated in terms of evolutive properties) is inapplicable.  In causal terms, exact space-time coordination breaks the causal thread ordinarily connecting the quantum object to its immediate past.
\end{remark}

\begin{consequence}[Non-deterministic outcomes]\label{cons:probabilism}
If an atomic position measurement is performed on a quantum object at~$t$, yielding outcome~$x$, and another such measurement is performed at~$t'$, yielding,~$x'$, then outcome $x'$ \emph{cannot} be determined by~$x$---the object's properties provide insufficient cause~(lacking an evolutive property conjugate to position) to account for its observed location.  Hence, in the absence of some external deterministic cause, the probability distribution over the possible outcomes at~$x'$ is expected to be non-deterministic.
\end{consequence}

\begin{consequence}[Effective Speed and Actual Speed]\label{cons:effective-speed}
If a classical particle is exactly located at~$x$ at~$t$, and is then found at~$x'$ at~$t' = t + dt$, then the cause of its change in its position can be attributed to its momentum at~$t$.  Hence, the quantity~$\bar{v}$ = $(x' - x)/(t' - t)$ is its \emph{actual} speed, \viz a measure of its \emph{motion} in~$[t, t']$.  However, this is not the case for an exactly located quantum particle.  In that case,~$\bar{v}$ is an \emph{an effective speed}---not `the speed of the particle', since implicit in this classical language is the assumption that the particle possesses an evolutive property which is a direct cause of its change of location. \end{consequence}

\begin{consequence}[Independence from speed of source]\label{cons:source-independence}
If a quantum particle is exactly located at~$x$ at time~$t$, it has no property that encodes the motion of the source which created it.  Hence, such a particle emitted from a moving classically-describable object~(the source) will exhibit an effective speed which is independent the speed of source. This provides a property-based understanding of Einstein's postulated independence of the speed of light from the speed of its source.  Given this correspondence, the so-called `speed of light' is an effective speed, not an actual speed~(of a photon).  
\end{consequence}

\begin{consequence}[Huygens' principle]\label{cons:Huygens}
If a quantum particle is exactly located at~$x$ at time~$t$, and is found at~$x'$ at~$t'$, then the cause of its change~(or non-change) in position must be attributed to factors extraneous to the object, such as the local space in which in is located at~$t$ and~$t'$.  By symmetry, one would expect the effective velocity~(see Consequence~\ref{cons:effective-speed}) to be isotropic relative to the local space.  Hence, taking into account Consequence~\ref{cons:probabilism}, \emph{a priori} one would expect an equiprobable likelihood of detection at~$t'$ on a sphere centered on~$x$, with the radius reflective of the structure of the local space.
\end{consequence}

\begin{consequence}[Speed of light]\label{cons:speed-of-light}
Continuing the example in Consequence~\ref{cons:Huygens}:~one would also expect the effective speed of the object to reflect the structure of the local space.  Since such a speed is fundamental, not reflecting any contribution from the object's properties, it can be used as a local probe of the geometry of space-time.  This speed corresponds to the one-way speed of light. 
\end{consequence}

\paragraph{Predictive Insufficiency.}  If an object is subjected solely to a sequence of atomic measurements of ICP~$P$, then the quantum closure postulate and the quantum property exclusivity postulate jointly imply that no predictive model can be constructed which can approximate deterministic change in~$P$~(see, in particular, Remark~\ref{rem:Bohr-coordination-causality}).  Hence, in order to construct models that can approximate the deterministic change in~$P$ which is characteristic of classical physics and reflective of ordinary experience of physical objects, we must expand consideration to \emph{non-atomic} measurements of~$P$.  Such an expansion, carried out below, requires engagement with the specifics of the reconstruction of Feynman's rules of quantum theory carried out within the operational framework.

\section{Complementarity of Quantum Properties}
\label{sec:complementarity}

As previously remarked, the process of operationalizing quantum properties in Sec.~\ref{sec:property-operationalization} is incomplete since, as we have seen, predictive insufficiency arises if we allow only a sequence of atomic measurement outcomes.  Although this scenario is the one typically considered in the idealized application of quantum theory, our approach to the operationalization of quantum properties requires that we explicitly consider non-atomic outcomes.

As we have seen in~Sec.~\ref{sec:property-attribution-non-atomic-outcomes}, property attribution following a non-atomic outcome~$\widetilde{m}$ is theory-dependent.  In classical physics, this attribution is made possible~(see Sec.~\ref{sec:classical-non-atomic-property-attribution}) through the assumptions of \begin{inlinelist} \item measurement passivity and \item the atomicity of property-values. \end{inlinelist}   However, due to the fact that quantum theory was not constructed on a systematically-constructed metaphysical foundation, it is not known whether these assumptions are appropriate in the quantum realm.  Hence, in order to discover how to attribute properties following a non-atomic outcome, we shall now examine the reconstruction of Feynman's rules.  From this reconstruction, we shall extract a precise property complementarity principle that specifies which properties can be attributed following a non-atomic outcome.

\subsection{Reconstruction of Feynman's rules}

In Sec.~\ref{sec:sequence-probabilities-and-interrelations}, we considered a relationship between sequences obtained from two different set-ups.  In one, the measurements~$\mment{L}, \mment{M}, \mment{N}$, are performed at times~$t_1, t_2, t_3$, yielding either~$\sseq{C} = \lseq{\ell}{m}{n}$ or~$\sseq{D} = \lseq{\ell}{m'}{n}$, where~$\ell, n$ are atomic, with~$m,m'$ possibly non-atomic and~$m \cap m' = \emptyset$.  The other set-up is similar, except~$\mment{M}$ in replaced by~$\mment{\widetilde{M}}$, and yields sequence~$\sseq{E} = \lseq{\ell}{\bubble{m}{m'}}{n}$. 

In the reconstruction of Feynman's rules~\cite{GKS-PRA}, the relationship between~$\prob{C}, \prob{D}$ and~$\prob{E}$ is \emph{not} obtained~(as in the classical case) by positing that measurements and properties have certain characteristics, but rather by representing each sequence with a real-number pair, and deriving a calculus of these pairs which is consistent with \begin{inlinelist} \item the so-called experimental logic~(formed by the sequence-combination operators,~$\pll, \ser$ and a unary so-called reciprocity operator); \item the probability-product equation, Eq.~\eqref{eqn:probability-product}; and \item the requirement that insertion of a so-called trivial measurement~(which has but a single possible outcome) does not interfere with repeatability. \end{inlinelist}

The resulting calculus can be rewritten so that each sequence,~$\sseq{A}$, is represented by a complex number~$\amp(A)$, which is additive over parallel combination.  This yields,
\begin{equation}
\label{eqn:amplitude-additivity}
\amp(E) = \amp(C) + \amp(D),
\end{equation}
which we recognize as Feynman's amplitude sum rule~\cite{Feynman48} in operational form.  The probability,~$\prob{A}$, associated with sequence~$\sseq{A}$ is given by
\begin{equation}
\label{eqn:probability-amplitude-equation}
\prob{A} = |\amp(A)|^2.
\end{equation}
Hence, writing~$\amp(C) = \sqrt{\prob{C}}\,e^{i\phi_1}$ and~$\amp(D) = \sqrt{\prob{D}}\,e^{i\phi_2}$,
\begin{equation}
\prob{E} = \prob{C} + \prob{D} + 2 \sqrt{\prob{C} \prob{D}}\,\cos(\phi_1 - \phi_2).
\end{equation}
This is the familiar `interference' characteristic of double-slit diffraction.

\subsection{Complementary Property Models}

As summarized above, the amplitude sum rule, Eq.~\eqref{eqn:amplitude-additivity}, has been reconstructed by positing that there exists a relationship between the sequences~($\sseq{C}, \sseq{D}, \sseq{E}$) obtained in two \emph{different} set-ups.  However, here we interpret the sum rule to mean that, in an experiment in which non-atomic outcome~$\widetilde{m}$ is obtained, there are two \emph{different} ways of conceptualizing the quantum object's properties immediately following the outcome:
\begin{itemize}
\item\emph{Non-contextual Property Model.} The object possesses a property with value~$\property{m''}$ where~$m'' \subset \widetilde{m}$ is some non-empty outcome that refines~$\widetilde{m}$.
\item\emph{Contextual Property Model.} The object possesses a property with non-atomic value~$\property{\widetilde{m}}$ corresponding to~$\widetilde{m}$.  
\end{itemize} 
The names of these models reflects their different perspective on the property concept.  The non-contextual property model asserts that obtaining outcome~$\widetilde{m}$ implies that the underlying property has, in fact, has value~$\property{m''}$ for some~$m'' \subseteq \widetilde{m}$, \emph{even though the outcome does not reveal that fact.}  Consequently, for either outcomes~$\bubble{m}{m'}$ or~$\bubble{m}{m''}$, the underlying property could be~$\property{m}$ even though the context~(both the specific measurement and its outcome) has changed.  In contrast, the contextual property model does not venture beyond the context---it simply posits property~$\property{\widetilde{m}}$ given outcome~$\widetilde{m}$.

\subsubsection{Connection to Bohr's wave-particle complementarity}
To better appreciate the novel physical content of these property models, suppose that a quantum object undergoes a non-atomic position measurement and that region-valued outcome~$R$ is obtained.  The non-contextual property model implies that the object can possess a point-valued position property,~$\vect{r} \in R$ immediately after outcome~$R$.  However, the contextual model attributes the position property~$R$ itself.  That is, according to the contextual model, the object is an \emph{extended simple}, \viz an object lacking proper parts but spatially extended.  Such objects play no established role in classical physics---even classical waves are reduced to a set of point-values~(\viz scalar or vector fields).  

This example establishes a close parallel to Bohr's wave-particle complementarity:~the non-contextual model roughly corresponds to the `particle' concept~(a continuously point-localized object), while the contextual model corresponds to the `wave' concept.  However, the above principle is wholly expressed in property terms, which considerably clarifies the ambiguities that arise due to Bohr's use of these loaded terms~(`particle', `wave').  In particular, from our current perspective, the `wave' concept represents \emph{an extended simple}, leaving out other aspects of the classical wave model~(such as the wave equation or wave parameters such as wavelength or speed).  Nonetheless, the parallel is not exact---there are important differences, most significantly that while Bohr's `particle' describes a point-localized entity, the non-contextual model allows for region-valued position properties in addition to point-valued ones.

\section{Actual and Potential Quantum Properties}
\label{sec:actual-potential-quantum-properties}

From the perspective of the reconstruction of Feynman's rules of quantum theory, the mathematical synthesis of the complementarity property models is key to the derivation of the Feynman amplitude sum rule.  Thus, the models are not separately `true', and should not be taken literally; rather, they are two different perspectives, each of which captures some fundamental truth of the situation, and which can be mathematically synthesized in spite of the fact that they make different assertions about the post-measurement properties of the system.

We can naturally combine the complementary property models using the notions of actual and potential as follows:
\begin{postulate}[Quantum Properties~(Single Individual)] \label{postulate:quantum-properties-single}
Immediately after a non-atomic measurement~$\widetilde{\mment{M}}$ has been performed on an individual object and non-atomic outcome~$\widetilde{m}$ obtained, the object has \emph{actual} property~$\property{\widetilde{m}}$, together with the set of \emph{potential} properties~$\property{m''}$ with~$m'' \subset \widetilde{m}$.
\end{postulate}
Here, the actual property directly reflects what is actually given to us in the present, namely the observed outcome~$\widetilde{m}$.  In contrast, the potential properties reflect the object's connectedness from the present into the immediate future, and express what \emph{could} be actualized in a subsequent measurement.  
The classical and proposed quantum model of properties are illustrated in Tables~\ref{tbl:classical-properties} and~\ref{tbl:quantum-properties}.

As we have seen, this model of quantum properties requires that properties are generalized from atomic to non-atomic valuedness.  In particular, position can be region-valued which, when applied to a physical object, describes an extended simple.   

The model of properties marks a sharp distinction between \emph{space-time} and the \emph{space of properties}:~space-time is the space of momentary outcomes~(or events), not the space of properties.  Rather, space-time functions as the \emph{base space} in terms of which properties ascribed to objects are defined.  This distinction is largely obscured in classical physics, where the primary time-dependent property, \viz position, of an object is simply a point in space-time.  But, in the proposed quantum property model, properties are considerably enriched:~\eg \begin{inlinelist} \item position is, in general, region-valued; \item such regions can be topologically disconnected; and \item there are both actual properties and potential properties. \end{inlinelist}

\setlength{\tabcolsep}{6pt}
\begingroup
\begin{table*}
\centering
	\begin{ruledtabular}
    \begin{tabular}{p{3.50cm}p{4.75cm}p{4.75cm}} 
    \smallskip
    {\bf } 															& \tableheadtext{Instantaneous Categorical Property} 			 & \tableheadtext{Evolutive Property}  \\ \hline

\smallskip\tableheadtext{Atomic Outcome,~$\vect{r}$}   	
																	& \smallskip\tabletext{Position~$\vect{r}$}   	
																	& \smallskip\tabletext{Evolutive property~$\vect{v}$} \\
    
\smallskip\tableheadtext{Non-atomic Outcome, $R$}  
																	& \smallskip\tabletext{Position~$\vect{r} \in R$}   
																	& \smallskip\tabletext{Evolutive property~$\vect{v}$}  \\   

\smallskip\tableheadtext{Remarks}   						
																	& \smallskip\tabletext{Registrativity implies position \newline attribution in non-atomic case.}
																	& \smallskip\tabletext{Evolutive attributions require \newline measurement passivity.}   \\

	\\
    \end{tabular}
	\end{ruledtabular}
\caption{\label{tbl:classical-properties} \emph{Classical properties.}  Classical property attributions following atomic and non-atomic position measurements. }
\end{table*}
\endgroup

\setlength{\tabcolsep}{6pt}
\begingroup
\begin{table*}
\centering
	\begin{ruledtabular}
    \begin{tabular}{p{3.50cm}p{4.75cm}p{5.75cm}} 
    \smallskip
    {\bf } 															& \tableheadtext{Instantaneous Categorical Property} 			 & \tableheadtext{Evolutive Property}  \\ \hline

\smallskip\tableheadtext{Atomic Outcome,~$\vect{r}$}   	
																	& \smallskip\tabletext{Position~$\vect{r}$}   	
																	& \smallskip\tabletext{No evolutive property} \\
    
\smallskip\tableheadtext{Non-atomic Outcome,~$R$}  
																	& \smallskip\tabletext{Position~$R$ \emph{actually};  \newline Position~$R' \subset R$ or~$\vect{r} \in R$ \emph{potentially}}   
																	& \smallskip\tabletext{Evolutive property conjugate to position \newline if position takes  non-atomic value}   \\

\smallskip\tableheadtext{Remarks}   						
																	& \smallskip\tabletext{}
																	& \smallskip\tabletext{Due to Quantum Closure and \newline Property Exclusivity}   \\

	\\
    \end{tabular}
	\end{ruledtabular}
\caption{\label{tbl:quantum-properties} \emph{Quantum properties.} Quantum property attributions following atomic and non-atomic position measurements. }
\end{table*}
\endgroup

\section{Applications}
\label{sec:examples}

In this section, we apply the model of quantum properties~(as proposed in Sec.~\ref{sec:actual-potential-quantum-properties}) to several examples.  These examples bring out different aspects and consequences of the model.

\subsection{Zeno's Arrow Paradox}
\label{sec:Zeno}

\subsubsection{Statement of the Paradox, and its Significance}
\label{sec:Zeno-statement}

In~\cite[\S3.5]{Lear-Aristotle}, Zeno's paradox of the arrow is formulated as follows:
\begin{itemize}
\item[]\textbf{Z1.}  \emph{Premise:} Anything that occupies a space just its own size is at rest.
\item[]\textbf{Z2.}  \emph{Premise:} A moving arrow, while it is moving, is moving in the present moment.
\item[]\textbf{Z3.} \emph{Premise: } But in the present, the arrow occupies a space just its own size.
\item[]\textbf{Z4.} \emph{Inference: }  Therefore, in the present, the arrow is at rest.
\item[]\textbf{Z5.} \emph{Inference: }  Therefore, a moving arrow, while moving, is at rest.
\end{itemize}
Zeno's arrow paradox was formulated during a period of time in which common notions such as location and time were being precisified and idealized.  The paradox brings into question whether the notion that an object `is moving' in the experienced now can be consistently formulated in such an idealized setting.  The essential point is that, if `the now' is idealized as a durationless instant~(analogous to a geometrical point), then a body appears~(in the mind's eye) the same~(`occupying a space just its own size') whether it is ostensibly moving or at rest.  But, if one accepts that motion could only ever occur in `the now', then motion itself seems impossible.

Below we briefly consider three standard attempts to resolve the paradox, before turning to a resolution based on the proposed model of quantum properties.

\subsubsection{Standard Resolutions of the Paradox}
\label{sec:Zeno-paradox-resolutions}

\paragraph{Classical mechanics: Velocity and Causality.}
Zeno's paradox presumes an absolute distinction between the arrow being at rest and in motion.  In view of Galileo's principle of relativity, this presumption is mistaken.  Accordingly, we take the nub of the paradox to be that, in an instant, if the arrow is \emph{exactly located}, it has no property that can represent its putative state of motion.  Hence, one cannot causally explain the arrow's location a moment later in terms of the properties it possesses in the present. 

Classical mechanics sidesteps the paradox by retaining the notion of durationless instant, but attributing additional properties to the arrow in an instant so as to restore causal explanation.  As discussed in Sec.~\ref{sec:operationalization-classical-properties}, this is achieved by ascribing both a position and a velocity to a body at an instant, thereby contradicting Z1.  However, it is unclear whether classical mechanics can be said to genuinely \emph{resolve} the paradox or whether it simply \emph{presumes}~(as an empirically self-evident given) that motion occurs and then \emph{describes} it.  The ambiguity here turns on the function of velocity---is it simply \emph{descriptive} of motion, or does in play a \emph{causal role} in accounting for the arrow's change of location via Newton's equations of motion~(see \eg~\cite{Lange2005})?

\paragraph{Russell:~Motion as Trajectory; Generalized Causality.}

The so-called `at-at' resolution posits that the only property of the arrow at an instant is its position~\cite[p.~473]{Russell1903}.  What we call \emph{motion} is then understood as merely a way of summarizing the fact that the arrow is at different locations at different instants of time.  Accordingly, unlike the situation in classical mechanics, a causal explanation of the arrow's motion cannot consist in the positing of a connection between the states of the arrow at two infinitesimally-close instants, but rather between the arrow's state (\emph{viz.} its location) at \emph{three} such instants. 

The main strength of this resolution is that it attributes only those properties to the arrow in an instant that are manifest in that instant~(\ie ICPs).  Its main weakness is that it gives up the common intuition that all of the causes of the arrow's present position pass---so to speak---through a given instant of time; or, to put it another way, that these causes are \emph{mediated} by the arrow's state at a single instant of time.

\paragraph{Bergson:~Rejection of `Cinematographic' Representation of Motion.}

According to Bergson, the arrow paradox rests on an erroneous view of time.  Specifically, in sharpening the everyday distinction between the past and the present to mean that, in the present moment, the past does not exist, one neglects the continuous~(or `flow') aspect of the passage of time.  Instead, one ought to regard the notions of past and present as fuzzy notions which, in some imprecise sense, \emph{interpenetrate}.  This view appears similar to the one espoused by Aristotle in his response to the paradox~(in~\cite{Pemberton2022}, this view is referred to as temporal holism). 

Responding to Zeno's arrow paradox, Bergson states: ``...the arrow never is at any point of its path. The most we can say is that the arrow could be at a certain point, in the sense that it passes through that point and would be free to stop there.''~\cite[p.~267]{Bergson2023}.   Here, Bergson makes a critical distinction between the arrow \emph{being} at a point of its path~(which he denies), and there existing the \emph{possibility} of it stopping such a point~(which he grants).  Thus, according to Bergson, the conception that the arrow occupies a sequence of positions at a succession of instants---a conception to which he refers as the \emph{cinematic} representation of motion---is a consequence of failing to make~(or collapsing) this distinction.  

The main limitation of Bergson's resolution is that it provides no positive indication on how one could faithfully describe motion in terms of the properties attributed to an object in a series of instantaneous snapshots.

\subsubsection{Resolution of the Paradox via Quantum Properties}

From the operational point of view, the most striking fact about Zeno's paradox is that it makes no mention of the concept of \emph{observation}.  Most of the previously proposed resolutions of the paradox~(such as those discussed in Sec.~\ref{sec:Zeno-paradox-resolutions}) are similar in this regard.  Rather, the paradox directly makes statements about the location and state of motion of an object.  Hence, the first step in applying the proposed model of quantum properties is to operationalize the situation.  

Such an operationalization presents an immediate choice:~should the act of observing the arrow's location be described as an atomic or as a non-atomic position measurement?  The former appears to be the best fit to the paradox as stated:~by Quantum Property Exclusivity, if an object has an exact position property, then it possesses no property associated with motion.  Hence, we at once reach a conclusion at odds with the intuition that the arrow is moving at an instant.

However, as previously discussed, a non-atomic position measurement is the proper representation of our act of observing where the arrow is in space. Since the measurement outcome is nonatomic, causal closure is \emph{not} established, so that the body retains some memory of its past; or, in causal terms, the causal thread connecting the present instant to infinitesimally prior instants remains intact.   If, then, a representative part of the arrow is observed in~$R$, it is located at~$R$ actually, and only potentially at a subregion~$R' \subset R$ or at a point~$\vect{r} \in R$.  In addition, it possesses an evolutive property which encodes its motion.  Hence, one can meaningfully say that the body \emph{is moving} in an instant.

Projected into space, a series of observations of the moving arrow generates a `trajectory' of sorts---a series of overlapping spatial regions, each corresponding to a region-valued measurement outcome.  Immediately after each such outcome, the arrow is located at such a region \emph{actually}.  At each instant, the arrow is \emph{not} at a point in space \emph{actually}, but only \emph{potentially}.  If it undergoes exact position measurement at some instant of time, it has the possibility of being found at a particular point within a finite region.  And, if such a measurement were performed, the body would have no evolutive property~(by Quantum Property Exclusivity).  This corresponds to Bergson's distinction between an arrow \emph{being} at a point~(we would say `\emph{potentially} being at a point'), and the \emph{possibility} that an arrow can be `stopped' at a point.

\emph{In summary:} The idealization of position measurement as atomic leads to a consequence closely resembling Zeno's premise~Z1, namely that nothing can be moving in an instant.  However, by representing position observations as non-atomic, and adopting a richer conception of properties drawn from quantum theory, it is possible to reflect the immediate past and anticipated future of the arrow in the properties attributed to it in the present moment.   We can thereby \emph{analyse} a temporally holistic process~(motion) into instantaneous slices without losing the essential process.

\subsection{Double Slit Diffraction}
\label{sec:double-slit}

Consider a double slit experiment in which an electron source is separated from a screen by a pair of slits.  Suppose that atomic position measurements show that an electron is emitted from point~$\ell$ and subsequently strikes point~$n$ on a screen.  At the slits, one can place a single detector which, given emission at~$\ell$ and detection at~$n$, always fires, yielding outcome~$\widetilde{m}$.   Alternatively, one can place two detectors~(\emph{which-way} detectors), one immediately before each slit, which always yields outcome~$m_1$ or~$m_2$ whenever~$n$ is subsequently registered.

Let us consider the first experiment.  Then, in terms of quantum properties, after outcome~$\widetilde{m}$ is obtained, the electron is at~$\widetilde{m} = \bubble{m_1}{m_2}$ actually, and is at~$m_1$ or~$m_2$ potentially.  The first property attribution reflects the electron's `wave' nature, the second its `particle' nature.   Hence, the electron behaves as a simple extended over a spatial region that encompasses both slits. Thus, its behaviour is expected to reflect the geometrical arrangement of both slits.  However, the electron also has the potential to be found at each slit under the appropriate conditions~(as is realized in the second experiment by employing which-way detectors).    

This qualitative property-based description closely tracks the formal operational-Feynman analysis of the experiment.  Such an analysis ascribes amplitudes~$z_1, z_2$ to the sequences~$\lseq{\ell}{m_1}{n}$ and~$\lseq{\ell}{m_2}{n}$, and thereby obtains the amplitude~$z = z_1 + z_2$ for sequence~$\lseq{\ell}{\widetilde{m}}{n}$.  At the formal level, the metaphysical meaning of these symbolic operations is opaque.  However, the quantum property model proposes a definite meaning in terms of actual and potential properties possessed by the electron.

The present example exhibits two important differences in comparison with the previous example of Zeno's arrow.  First, although an object is an extended simple~(actually) in both cases, the extended region is here \emph{topologically disconnected}.  This departs from the intuition that a `coarse' observation is described by a simply connected region of space, and here is made possible through an engineered arrangement of slits.   Second, the present example relies on the probabilistic element of detectors at the screen to reveal the extended nature of the electron.

\subsection{Entanglement}

Consider a system consisting of two non-identical subsystems~$\system{A}$ and~$\system{B}$, which undergo exact position measurements at~$t_1$ and~$t_3$, and non-atomic \emph{joint} position measurements at~$t_2$.  For simplicity, suppose that, if an atomic measurement were performed on either system at~$t_2$, it would yield only one of two possible atomic outcomes.
The composite sequence is~$C = \lseq{\ell}{\widetilde{m}}{n}$, where~$\ell, \widetilde{m}, n$ symbolize the composite outcomes:
\begin{align}
\ell  &= (\ell^\system{A}, \ell^\system{B}) \\  
\widetilde{m} & \subseteq \{(m_1^\system{A}, m_1^\system{B}), (m_1^\system{A}, m_2^\system{B}), (m_2^\system{A}, m_1^\system{B}), (m_2^\system{A}, m_2^\system{B}) \} \\
n &= (n^\system{A}, n^\system{B}).
\end{align}
The amplitude of the composite sequence~$C = \lseq{\ell}{\widetilde{m}}{n}$ is given by the amplitude sum rule as
\begin{equation}
\label{eqn:composite-amplitude-general}
\amp \bigl( \lseq{\ell}{\widetilde{m}}{n} \bigr) = \sum_{k \in \widetilde{m}} \amp \bigl (\lseq{\ell}{k}{n} \bigr).
\end{equation}
In the special case where the systems do not interact in~$(t_1, t_3)$, the amplitude~$\amp \bigl( \lseq{\ell}{\widetilde{m}}{n} \bigr)$ can be computed via the so-called \emph{composite system rule}~\cite[\S5]{Goyal2014},~$\amp(\sseq{A} \times \sseq{B}) = \amp(\sseq{A}) \cdot \amp(\sseq{B})$, where~$A, B$ are sequences for the outcomes of systems~$\system{A}, \system{B}$.  Hence,
\begin{equation}
\label{eqn:composite-amplitude-noninteracting}
\amp \bigl( \lseq{\ell}{(m_i^\system{A}, m_j^\system{B})}{n} \bigr) = \amp \bigl( \lseq{\ell^\system{A}}{m_i^\system{A}}{n^\system{A}} \bigr) \cdot \amp \bigl( \lseq{\ell^\system{B}}{m_j^\system{B}}{n^\system{B}} \bigr) \quad \quad \text{(no interaction)}
\end{equation}
As shown in~\cite{Goyal2014}, this is the amplitude form of the standard von Neumann tensor product rule.

Our quantum property model can be straightforwardly generalized to accommodate the structure of Eq.~\eqref{eqn:composite-amplitude-general} as follows:
\begin{postulate}[Quantum Properties~(Composite system)] \label{postulate:quantum-properties-many}
Suppose that a non-atomic measurement is performed on a composite system with non-identical subsystems~$\system{A}, \system{B}$, and yields non-atomic outcome~$\widetilde{m}$.  Immediately following the outcome, the system has property-value~$\property{\widetilde{m}}$ corresponding to~$\widetilde{m}$ actually, and property-value~$\property{k}$ with~$k \subset \widetilde{m}$ potentially.
\end{postulate}
For example, if particles~$A$ and~$B$ undergo a joint position measurement that yields non-atomic outcome~$R \subseteq \numberfield{R}^3 \times \numberfield{R}^3$, then they are located at~$R$ actually, but potentially \eg at any point~$(\vect{r}^\system{A}, \vect{r}^\system{B}) \in R$.  In case~$R = \{(\vect{r}_1,  \vect{r}_2), (\vect{r}_2,  \vect{r}_1) \}$, the particles are extended over a topologically-disconnected region~$R$, but potentially occupy two different configurations, \viz $(\vect{r}_1,  \vect{r}_2)$ or~$(\vect{r}_2,  \vect{r}_1)$.  This contrasts with classical property attribution, which would in this case assert on the basis of the non-atomic outcome that particles~$A, B$ are \emph{in fact} located at configuration-space point~$(\vect{r}_1,  \vect{r}_2)$ \emph{or} $(\vect{r}_2,  \vect{r}_1)$.

On the basis of this extended quantum property model, the set of possible correlations between possible outcomes at~$t_3$ is conceivably enriched relative to the classical property model.  From what has been shown above, it is not evident that these correlations are detectable; but, in fact, we know that they are revealed through Bell-inequality violation.  In terms of properties, we can understand such violation as due to the fact that the composite system is an \emph{extended complex} actually---a system of two individuals that is extended over~$\numberfield{R}^3 \times \numberfield{R}^3$.  Hence, in comparison with the double-slit diffraction example~(Sec.~\ref{sec:double-slit}), where the phenomenon of diffraction is due to an extended simple, Bell-inequality violation is due to an extended complex.

\section{Comparison with Related Work}
\label{sec:related-work}

As discussed in Sec.~\ref{sec:properties}, the concept of property is fundamental to our everyday conception of physical objects.  From that domain, it is carried over in precisified form into classical physics, where key distinctions~(such as between time-independent and time-dependent properties) and assumptions~(such as the primacy of spatial over non-spatial properties) are made.  Our sense of \emph{understanding} the meaning of classical physics rests, in large part, on such conceptual continuities.  Accordingly, ascertaining the appropriate way to apply the property concept to quantum theory is one way to better understand the theory, in particular by understanding how the reality described by quantum theory resembles, or differs from, that described by classical physics.  In this paper, we have taken Bohr's coordination--causality complementarity as the starting point for the development of a property model fitted to quantum theory.  In this section, we briefly discuss some of the many other attempts to apply the property concept to quantum theory, and contrast both their methodology and assertions with those in the present work.

\subsection{The Eigenstate--Property Posit and its Limitations}
\label{sec:eigenstate-property-posit}

It is commonly asserted that if a quantum system is in an eigenstate,~$\ket{a}$, of a measurement operator,~$\op{A}$, then it possesses a property-value corresponding to the property measured by measurement procedure~$\mment{A}$ described by~$\op{A}$.  This \emph{eigenstate--property posit} rests on two ideas.  First, that since we can predict with certainty that performing the measurement described by~$\op{A}$ upon the system would yield outcome-value~$a$ with certainty, we have operational grounds for ascribing a \emph{property} and a \emph{property-value} to the system.  Secondly, that one can view~$\op{A}$ as \emph{representing} this property.

Since every pure state~$\ket{v}$ is an eigenstate of some Hermitian operator,~$\op{V}$, it follows from the eigenstate--property posit that a quantum system in a pure state \emph{always} possesses a definite property-value corresponding to some measurement operator.   This reflects the basic fact that, in abstract quantum formalism, all Hermitian operators are on the same footing~(in the sense that there exists a similarity transformation that takes any given Hermitian operator to any other).

However, there are a number of reasons to question the eigenstate--property posit.
First, as discussed in Sec.~\ref{sec:quantum-properties-introduction}, position measurements are granted primacy in the application of the quantum formalism to construct models of particles.  That is, idealized \emph{classical} position measurement procedures are assumed to constitute idealized quantum position measurement procedures, an assumption that is \emph{not} extended to classical properties such as velocity or momentum.   Operators corresponding to momentum measurements are then determined through an indirect, \emph{formal} process,  resulting in a quantum momentum measurement procedure which bears no resemblance to the classical momentum measuring procedure.   Hence, the assertion that `a particle has a definite value of momentum', made on the basis of the system being in a momentum eigenstate, \emph{does not have the same operational meaning} as in the classical context. 

Second, if one grants the eigenstate--property posit, then property-types are infinite in number, and most have no classical counterpart.  For example, if a structureless quantum particle is in state~$\psi(\vect{r})$ that is not an eigenstate of position or momentum, the particle possesses~(according to the eigenstate--property posit) a definite value of a property, but this property has no classical counterpart.  Hence, the attribution of the property concept in this case has very limited explanatory value.

Third, the eigenstate--property posit does not distinguish between non-degenerate and degenerate operators.  However, from an operational standpoint, this distinction is highly relevant since one only \emph{knows} that a system is in a pure state if it has been prepared via a measurement described by a non-degenerate operator.  However, practically realizable measurements are only sometimes describable via non-degenerate operators.  For example, realizable position measurements are \emph{never} so describable.  Thus, from an operational standpoint, it is vital to consider degenerate measurement operators.    Now, as we have seen in the discussion of causal closure, a non-atomic outcome of measurement~$\mment{A}$, which is described by a degenerate operator~$\op{A}$, does not establish closure with respect to a subsequent measurement~$\mment{A}$.  In contrast, if~$\op{A}$ is non-degenerate, all of its outcomes are atomic, any of which establishes closure.  In the property model we have developed, these two situations yield radically different property assignments:~after an atomic outcome, the system possesses only an actual property corresponding to~$\mment{A}$~(if this is a measurement of an ICP); but, after a non-atomic outcome, the system additionally possesses a potential property~(see Table~\ref{tbl:quantum-properties}).

In short, although the eigenstate--property posit seems \emph{prima facie} reasonable, its value in understanding the relationship between quantum and classical physics is limited.  As mentioned in the Introduction, these limitations stem from the underlying methodology, basing metaphysical analysis wholly on the abstract quantum formalism, neglecting \begin{inlinelist} \item the physical meaning of the quantum formalism's mathematical structures; and \item the fundamental assumptions employed during its application. \end{inlinelist}  In contrast, the model developed in the present work acknowledges the primacy accorded to ICPs, such as position, in both everyday perception and classical physics, and takes into account at the outset the fact that ICPs such as position are accorded special status in the application of the abstract quantum formalism to build quantum models of specific systems.   It adopts an operational stance on property assignments, and uses casual analysis to systematically establish the basis of property assignments~(of both categorical and non-categorical properties).  And finally it harnesses a reconstruction of the quantum formalism to extract a detailed property-model which provides reliable intuitions about key scenarios.

\subsection{Potentiality and Actuality in Quantum Theory}
 
If a quantum system is \emph{not} in an eigenstate of a given measurement operator~(which is the generic situation), then the above-mentioned eigenstate--property posit does not apply.  In this case, the quantum state determines the outcome probabilities of the measurement that corresponds to the given operator.  The question then arises how to appropriately apply the property concept in such an instance.   Margenau and Heisenberg in the mid-1950s~\cite{Margenau1954, Heisenberg1955, Heisenberg1958} proposed that the evolving state be considered an objective, potential property.  For example, according to Heisenberg~\cite[pp.~52--53]{Heisenberg1958}:
\begin{quote}
``Now, the theoretical interpretation of an experiment starts with the two steps that have been discussed. In the first step we have to describe the arrangement of the experiment, eventually combined with a first observation, in terms of classical physics and translate this description into a probability function. This probability function follows the laws of quantum theory, and its change in the course of time, which is continuous, can be calculated from the initial conditions; this is the second step. The probability function combines objective and subjective elements. It contains statements about possibilities or better tendencies (``potentia'' in Aristotelian philosophy), and these statements are completely objective, they do not depend on any observer; and it contains statements about our knowledge of the system, which of course are subjective in so far as they may be different for different observers. In ideal cases the subjective element in the probability function may be practically negligible as compared with the objective one.''
\end{quote}

This line of thinking has been carried forward by several other authors~(\eg~\cite{Aerts2010,Heelan2016, Shimony1997, Suarez2004,Karakostas2007, Jaeger2017, Kastner2018}), who employ a range of closely related terms, \viz potentiality, latency, and propensity~(see~\cite{Suarez2007} for a review of the proposals of Margenau, Heisenberg, Maxwell, and Su\'arez).  The common thread in these interpretations is the posit that the quantum state consists of objective potentialities, with the notion of actuality ascribed either to measurement outcomes or to the property-values they are presumed~(via the eigenstate--property posit) to correspond to. In particular, the wavefunction is said to consist of a set of evolving potentialities, which are then quantified as probabilities via the Born rule.  

Most of this body of work shares the limitations described in Sec.~\ref{sec:eigenstate-property-posit}, namely exclusive focus on the abstract quantum formalism, to the exclusion of modelling heuristics and operational considerations.  Of particular interest in this connection is the work of Kastner et. al.~\cite{Kastner2018}, which, in contrast to prior work in this area, utilizes the Feynman formulation of quantum theory~(rather that the standard Dirac--von Neumann state formulation), and also emphasises the primacy enjoyed by position by virtue of its observabillity:
\begin{quote}
``However, the de Broglie waves corresponding to quantum states are not spacetime objects; it is only discrete, localized phenomena that are in-principle- observable elements of spacetime. In terms of our non-substance dualism, the de Broglie waves are the possibilities (\emph{res potentia}), while the discrete localized phenomena are the actualities (\emph{res extensa}).''
\end{quote}

The present work shares with the above-mentioned authors the broad goal of developing a realist view of quantum states, and the willingness to consider both categorical and non-categorical properties as objective existents.  However, it differs from previous work in several respects:
\begin{enumerate} 
\item The quantum property model developed herein is constructed from the ground up within an \emph{operational} framework that is sufficiently rich as to accommodate a reconstruction of quantum theory.  This framework brings into focus at the outset the connection between causality and property-attribution, which is an overarching theme in this work.  It also makes explicit how property attributions are based on specific assumptions concerning (a)~which property measurements establish casual closure and~(b) the characteristics of measurement~(such as registrativity, passivity).  The connection between causality and property establishes a direct conceptual linkage to Bohr's coordination--causality complementarity principle.
 
\item Previous work engages with the completed formalism of quantum theory directly.  The main disadvantage of such a methodological choice is that the formalism is couched in abstract mathematical terms, which obscures the connection between the mathematical structures and elementary concepts~\cite{Goyal2025a,Goyal2025d}.  For this reason, a few authors, such as~\cite{Kastner2018}, focus on the Feynman formulation, which we agree is advantageous in view of the vividness with which its core ideas can be displayed.  However, to overcome the obscuration inherent in any approach that directly interprets a completed formalism, the core ideas proposed here are extracted from postulates used in an operational reconstruction of Feynman's formulation of quantum theory.  As these postulates are formulated in operational terms, the connection between the proposed model and the operational facts and minimal assumptions needed to build predictive models within that operational framework is explicit. 

\item Rather than applying the terms actual and potential to different types of entity~(\eg measurement outcome and quantum state), as is usually the case, we apply these terms directly to properties that a quantum system can possess, and these properties are directly based on ICPs.  The contrast between classical and quantum physics is thereby brought sharply into focus, since they can be compared in how they attribute properties on the basis of the same set of observations.
\end{enumerate}

\section{Discussion}
\label{sec:discussion}

We conclude with a brief summary of the main ideas and results described herein, followed by brief reflections, and a few open questions. 

\subsection{Summary}

In this paper, we have explored the premise that the fundamental point of departure of quantum theory from classical mechanics is a fundamentally altered relationship between the concepts of causality and property.  In order to systematically elucidate this relationship, we have adopted an operational standpoint, according to which properties are not pre-given facts of the matter, but are attributed to objects in the process of constructing a predictive model.  Hence, properties and causality are inextricably connected.

In the operational setting, the primary difference between classical and quantum physics is that they are governed by different causal closure conditions.  The Quantum Closure postulate asserts that atomic measurement of a single instantaneous categorical property, such as position, establishes closure.  In the operational framework, this directly accounts for the loss of evolutive properties and the indeterminacy of subsequent ICP measurements.  Although not integral to the mainline of the paper, we have also noted that this may provide an understanding of other well-known physical postulates, such as Einstein's postulate of the independence of the one-way speed of light from the source, and the use of light as a fundamental measure of spacetime.

By harnessing a reconstruction of Feynman's rules of quantum theory, we have formulated a model of quantum properties that determines the properties of a system immediately following a non-atomic measurement outcome.  This model departs from the classical mechanical property model in two key respects: \begin{inlinelist} \item properties can be non-atomic; and \item properties can be potential as well as actual. \end{inlinelist}   The first implies that, in the case of position, a single individual can be spatially extended, and thus `wave-like' in Bohr's sense; and that a system of non-identical individuals can be an extended complex, and thereby be capable of exhibiting non-local behaviour.  The second implies that properties are of two fundamental types---\begin{inlinelist} \item those that directly reflect what is observed in the present; and \item  those that reflect connectivity of the present with the immediate past and future. \end{inlinelist}  These are equally fundamental and necessary to describe change through snapshots.

We have shown that this quantum property model provides a straightforward resolution of Zeno's paradox~(which resolution can be regarded as an elaboration and precisification of the informal resolution proposed by Bergson), and provides an intuitive way of understanding such characteristic quantum phenomena as double-slit diffraction and entanglement~(through the notions of extended simple and extended complex, respectively).

\subsection{Reflections}

Quantum theory departs from classical physics in several different ways, including indeterminacy, holistic behaviour, and entanglement.  The central challenge in grasping the theory is to find a core idea that somehow makes it clear why all of these departures occur in one go, and how they follow from that key idea.  From what has been described in this paper, the core idea is simply that, in the quantum realm, closure is established by atomic measurement of a single ICP~(not \emph{two} such measurements as in classical mechanics).  This leads to a radical indeterminism.  Non-atomic measurements moderate this indeterminism, which leads to the proposed model of quantum properties.  This model naturally describes individuals as \emph{extended entities}, which accounts for holism at the individual level and the possibility of non-local behaviour at the composite system level. 

To return to our starting point, namely Bohr's coordination--causality complementarity:~the essential challenge in understanding change is that it necessarily occurs over a period of time, yet we seek to understand change through a series of snapshots.  As Zeno's arrow paradox made clear long ago, there is a fundamental tension between these two perspectives.  We contend that classical mechanics side-steps, rather than resolves, this tension, and that it surfaces again in the quantum realm.  What emerges through our analysis is that non-atomic observations, and correspondingly non-atomic properties, are key to charting a middle-way through these perspectives---a non-atomic observation forces a process to leave a trace in the present without completely disrupting its temporally holistic nature.   Remarkably, non-atomic properties, introduced to address the problem of grasping change, lead one to expect holistic behaviour of systems of elementary objects.  Hence, it appears that the temporally holistic nature of an object~(and the change it undergoes) is inextricably connected to its spatial holistic behaviour.

\subsection{Open questions}

\begin{enumerate}
\item\emph{General derivation of quantum property model.} The induction of the model of quantum properties proposed herein ultimately relies on a specific reconstruction of quantum theory.  Is it possible to derive such a property model without recourse to a full reconstruction of quantum theory?  More pointedly:~what are the fewest operationally-formalizable conditions which are sufficient to obtain such a property model?  The framework of operational probabilistic theories or generalized probabilistic theories, which have been extensively used to investigate the interrelations between various characteristics of quantum theory~(such as entanglement and local tomography~\cite{Barrett2006}) would seem to be the most likely candidates in which such a question could be systematically answered.

\item\emph{Properties in systems of identical particles.} The quantum property model proposed herein is restricted to composite systems of non-identical particles.  However, a recent reconstruction of the quantum rules for handling systems of identical particles~\cite{Goyal2015} has  been interpreted~\cite{Goyal2019a,Goyal2023a} to mean that these rules arise from a new principle of complementarity~(the Complementarity of Persistence and Nonpersistence).  According to this principle, in the modelling of detection-events arising from measurements on such a system, one needs to consider two different object-models, \viz \begin{inlinelist} \item there are several individual particles~(the persistence model) and \item  there is a single holistic entity~(the nonpersistence model). \end{inlinelist}  How can the proposed property model be generalized to apply to such a system, whilst taking into account these two object-models?
\end{enumerate}

\newpage

\bibliographystyle{plainnat}

\bibliography{references_master}

\end{document}